Multimodal Radiomics Model for Predicting Gold Nanoparticles Accumulation in Mouse Tumors

Jiajia Tang[1], Jie Zhang[1], Jiulou Zhang[1], Yuxia Tang[1], Hao Ni[2], Shouju Wang[1]

1. Laboratory of Molecular Imaging, Department of Radiology, The First Affiliated Hospital of Nanjing Medical University, Nanjing, Jiangsu, China
2. Department of Mathematics, University College London, London, UK

1. Introduction

Preclinical research has demonstrated that nanoparticles (NPs) can effectively accumulate in solid tumors as diagnostic and therapeutic agents for cancer. This phenomenon can reduce side effects and increase precision known as the enhanced permeability and retention (EPR) effect. However, only a small portion of NPs have successfully transitioned from the lab to the clinic. The primary obstacle is that NPs delivery varies widely among NPs and patients, leading to poor treatment effects in patients lacking adequate delivery.

Lin et, al. (2023) reported that the median tumor delivery efficiency (DE) following intravenous (iv) administration spanned from 0.81% ID/g (NPs in gliomas) to 11.45% ID/g (dendrimer NPs) (above 14 folds change) by analyzing the literature data published from 2005 to 2021. The average therapeutic efficacy demonstrated in an unselected patient cohort fails to surpass that of comparable small molecule therapies, contributing to the disappointing attrition rate observed in nanomedicine clinical trials. Parser of nanomedicine tumor delivery patterns at the in vivo level, and subsequently identifying individuals with heightened NPs accumulation are crucial for optimized therapeutic benefits and maximizing the potential of NPs. However, the tumor delivery patterns of nanomedicines are far more complex compared to small molecules. This complexity stems from the multifaceted NPs with variable physicochemical properties and the pronounced tumor heterogeneity, which present significant challenges in patient stratification. NPs, with diverse sizes, potentials, morphologies, and deformabilities, exhibit distinct fates in vivo. Tumors, meanwhile, manifest remarkable heterogeneity not only inter-tumoral, across distinct malignancies, but also intra-tumoral, within individuals harboring the same tumor type. This heterogeneity encompasses blood perfusion dynamics, tissue stiffness, architectural organization, and more, exhibiting considerable inter- and intra-patient variability but decisive in governing nanoparticle accessibility, ultimately influencing the success of nanomedicine-based interventions.

To address these intricacies, research have turned to machine learning (ML) algorithms as powerful tools to decipher the multidimensional landscape of nanomedicine tumor delivery. ML models can assimilate and analyze a vast array of data points, including nanoparticle characteristics (such as size), tumor biological features (like expression of specific receptors, microenvironmental factors), and patient-specific parameters (genetic profiles, demographic information), thereby uncovering subtle but crucial correlations and predictive patterns. By leveraging these algorithms, it becomes possible to parse through the high-dimensional data and identify key determinants that influence the variability in nanoparticle delivery.

In the clinical setting, medical imaging is one of the most commonly used screenings for cancer patients and is a non-invasive technique. Radiomics evolving from imaging, emerged as powerful tools for in vivo assessment of both nanoparticle behavior and tumor status, offering a quantitative and comprehensive perspective beyond the limitations of traditional visual interpretation. Combining ML and radiomics allows for assessing the relative importance of features derived from NPs and tumors, meanwhile, confirming the significance of imaging data in predicting delivery to solid tumors.

In this study, seven tumor xenograft models were established, and nanoparticles of three varying sizes were synthesized to investigate the differences in gold nanoparticles accumulation in tumors by integrating radiomics into ML algorithms. This strategy may not only accelerate the development process but also increase the likelihood of clinical success by personalizing treatment plans to maximize therapeutic outcomes.

## 2. MATERIALS AND METHODS

### 2.1. Synthesis and Characterization of PEGylated Gold Nanoparticles

All glassware was washed with soap, aqua regia (hydrochloric acid: nitric acid = 3 : 1), and deionized water before use. GNPs with 15 nm diameters were synthesized according to the Frens/Turkevich method, whereby ionic gold was reduced into solid gold via citrate at high temperatures. Sodium (1 mL of 3% (w/v)) citrate tribasic dehydrate (Sigma S4641) and chloroauric acid tetrahydrate (25 μL 1M, Sigma 254169) was added to 99 mL of boiling deionized water under vigorous stirring. This mixture solution was boiled and stirred for 7 min, producing a wine-colored solution, and then immediately cooled in an ice water bath. GNPs with larger sizes (40 and 70 nm) were grown on 15 nm particles. In detail, the reaction flask containing 8.5 mL (40nm) or 9.75 mL (70nm) deionized water, 94.75 μL 250 mM chloroauric acid tetrahydrate, 96.29 μL of 4.41% (w/v) (150 mM) sodium citrate tribasic dehydrate, 1.5 mL (40 nm) or 0.25 mL (70 nm) of 5 nM 15 nm particles and 96.29 μL of 2.75% (w/v) (250 mM) hydroquinone (Sigma H17902), stand overnight at 4°C. Washing proceeded as follows: centrifugation at 7000 g (15 nm), 3500 g (40 nm), or 1000 g (70 nm) for 45 min at room temperature, to remove reaction by-products. The stock solutions were stored at 4°C in the dark until the surface modification was performed.

Particles were PEGylated by mixing 2 mL 10 nM mPEG2K-SH and 9 mL 10 nM (15 nm), 1 nM (40 nm), or 0.5 nM (70nm) GNPs, incubating at 60°C for 1h. PEGylated Gold Nanoparticles were washed twice in Eppendorf tubes by centrifugation at 12000 g (15 nm PEG-GNPs), 3000 g (40 nm PEG-GNPs), 1000 g (70 nm PEG-GNPs) for 60 min at 4°C. The supernatant was removed and the pellet was resuspended in 0.5 mL of water.

Transmission Electron Microscopy (TEM, JEOL JEM-2100, Japan) was used to characterize the average diameter and morphology of gold nanoparticle cores. The dynamic light scattering (DLS) and zeta potentials of the synthesized GNPs before and after surface modification were determined by dynamic light scattering measurement Brookhaven analyzer (Brookhaven Instruments Co., Holtsville, USA). The Ultraviolet-Visible (UV-Vis) measurements were analyzed by UV–Vis spectrophotometer (UV-3600, Shimadzu), which was used to determine the extinction spectrum of each nanoparticle dispersion and to estimate the nanoparticle molar concentrations.

### 2.2 Cell Culture and Tumor Model

All animal research was reviewed by and conducted according to the Animal Core Facility of Nanjing Medical University (protocol number 2210046). The six-week-old female C57BL/6 mice were purchased from GemPharmatech Co., Ltd. Seven tumor subcutaneous xenograft models were successfully constructed as follows: B16 (Melanoma), E0771 (Breast Cancer), Hepa1-6 (Hepatocarcinoma), LLC (Lewis Lung Carcinoma), MB49 (Bladder Carcinoma), MC38 (Colon Carcinoma), PAN02(Pancreatic Adenocarcinoma). Approximately $1\sim2 \times 10^6$ cells of each tumor were subcutaneously injected into the flanks of the mice. When the tumors reached approximately 500 mm$^3$, the mice were used for the following experiments.

2.3 Assessing the Distribution of PEG-GNPs in Tumors

When the CT and ultrasound image acquisition were completed, mice were intravenously injected with 15 nm PEG-GNPs, 40 nm PEG-GNPs, or 70 nm PEG-GNPs. The injection dose of PEG-GNPs was approximately 1mg and the exact dose was determined by the inductively coupled plasma optical emission spectrometry (ICP-OES). After 24 h, the tumors of mice were harvested and weighed. Then, the tumors and injection dose solutions were digested to estimate the Au mass by ICP-OES. The Au quantity contained in tumors was expressed as percentage of injected dose per gram of tissue (%ID/g) according to the following formula: %ID/g = $Au_{tumor}/Au_{injection} \times G_{tumor} \times 100\%$, where $Au_{tumor}$ is mass of Au in tumors, $Au_{injection}$ is mass of Au in injection dose, $G_{tumor}$ is the mass of tumor.

2.4. Image Acquisition and Analysis

The computed tomography (CT) images of tumors were scanned in the prone position with a dual-source CT system (Somatom Force, Siemens Healthcare, Forchheim, Germany). The scanning parameters were shown as follows: slice thickness of 1 mm, tube voltage of 120 kVp, and tube current-time product of 100 mAs. The CT values and volumes of the tumors were subsequently quantified. The CT value was determined by averaging three consecutive measurements taken at the three largest levels of the tumor. The tumor volume was calculated on CT images according to the following formula: $V = \pi/6 \times L \times W \times H$, where L, W, and H represent the longitudinal diameter, width, and height of the tumor, respectively.

The ultrasound equipment used was Philips EPIQ elite Ultrasound system (Philips Healthcare, Washington, USA) with Philips EL18-4 probe. Ultrasound image types include B-mode US(BMUS), shear wave elastography (SWE), and contrast-enhanced ultrasound (CEUS). The maximum diameter plane was selected based on the US images of each tumor. To obtain stable and standard shear wave elastography images, the probe should be kept still for about 5-10 seconds per scan and avoid pressing the tumor. A representative SWE image was selected for the calculation of Young's modulus of the tumor (SWE-mean) (kPa). Then the mice were intravenously injected with 0.2 mL 5 mL/mg of microbubbles containing sulfur hexafluoride (SonoVue, Bracco, Milan, Italy). The real-time CEUS examination was recorded immediately after contrast injection and lasted 90 seconds. Those videos were stored for quantitative analysis with Philips built-in analysis software (QLAB 9.0™ Philips Medical System, Washington, USA). For each tumor, the video spanning from 1 s to 90 s was taken to perform the perfusion time-intensity curve (TIC) analysis. The region of interest (ROIs) was plotted along the boundary of each tumor. Then the parameters of the time intensity curve were obtained as follows, wash-in slope (WIS), peak intensity (PI), time to peak (TTP), and area under the curve (AUC). A representative image was saved at the peak time for tumor segmentation.

2.5 Tumor Segmentation and Radiomics Feature Extraction

ROIs of tumors were manually delineated in three imaging modalities, where CT images delineated were full slices of the tumor and US images delineated were maximum slices of the tumor, using the open-source software ITK-SNAP 3.8.0 (http://www.itksnap.org). We used the open-source Pyradiomics package (version: 2.12; https://pyradiomics.readthedocs.io/en/2.1.2/) to extract an extensive range of radiomics features from each tumor ROI, including first-order statistics, shape features (2D, 3D), textual features, and wavelet features. Accounting for multimodal radiomics from the CT and US images, a total of 3299 features were obtained per tumor, including 1389 CT radiomics features, 955 B-mode features, and 955 CEUS features. The feature algorithms can also be found online (https://pyradiomics.readthedocs.io/en/latest/). The calculated feature values were standardized with Z-scores. The Z-Score calculation formula is: $Z = \frac{X-\mu}{\sigma}$, where $X$ is the value to be calculated, $\mu$ is the mean of the feature values, and $\sigma$ is the standard deviation of the feature values.

2.6 Feature Processing and Selection

To assess intra-observer consistency, 37 tumors were randomly selected from 183 tumors for delineation twice, with a one-month interval between the two ROI delineations. Radiomics features with an ICC (Intraclass Correlation Coefficient) > 0.80 were included in the subsequent feature selection, totaling 2560 radiomics features, including 791 CT features, 870 B-mode grayscale features, and 899 CEUS features.

A total of 183 tumors were randomly divided into a training set and a test set, with the training set consisting of 121 tumors and the test set of 62 tumors. The median GNP accumulation in the training set was used to divide all tumors into high and low uptake groups. The PCC (Pearson Correlation Coefficient) method was used, setting a threshold of 0.95 to identify and address feature pairs with correlations of 0.95 or higher, reducing redundancy in the feature set to retain features that provide independent information.

In the classification model, the LASSO (Least Absolute Shrinkage and Selection Operator) regression algorithm was utilized to select meaningful features and eliminate unimportant ones through variable selection and regularization, enforcing model sparsity via the λ coefficient. Radscores for CT_radscore and BMUS_radscore were calculated based on the selected features and coefficients for subsequent model construction. The specific calculation formulas are provided in Supplementary Materials.

In the regression model, feature selection was initially conducted using Mutual Information (MI), which measures the extent of information sharing between features and the GNP accumulation. This method helps identify the features that are most strongly associated with the target variable. The MI score for each feature reflects its contribution to the prediction of the target variable, and based on these scores, we select the features with the highest information content. Subsequently, the K-Nearest Neighbors (KNN) algorithm is used to predict the GNP accumulation using these selected features.

In terms of tumor biological features, gold nanoparticle features, and quantitative imaging features, tumor type, the diameter of the injected gold nanoparticles, and curve types (wash-out, plateau curve) are categorical variables and were converted into binary (0 or 1) features using One-Hot Encoding. Tumor volume, average tumor CT value, mean SWE, perfusion slope (WIS), peak intensity (PI), time-to-peak (TTP), and area under the curve (AUC) are continuous variables. The preprocessing method used was MinMax Scaling, which scales values ranging between 0 and 1. The formula for this scaling is $X_{scaled} = \frac{X - X_{min}}{X_{max} - X_{min}}$, where X is the original data, $X_{min}$ is the minimum value, and $X_{max}$ is the maximum value in the data. Each of the aforementioned features was individually included in a univariate logistic regression analysis to identify those that are valuable for predicting the accumulation of gold nanoparticles (P < 0.05). The features selected from the univariate logistic regression analysis were then included in a multivariate logistic regression analysis. Features that were found to be predictive in both univariate and multivariate analyses were subsequently incorporated into the model construction.

2.7. Development and performance of model

The radiomics composite model was developed by combining tumor category, tumor volume, the diameter of injected gold nanoparticles, average tumor CT value, mean SWE value, and TIC curve parameters including curve type (wash-out, plateau curve), wash-in slope (WIS), peak intensity (PI), time to peak (TTP), and area under the curve (AUC) with CT radscore, BMUS radscore, and CEUS radscore. The baseline model consists of tumor category, and the diameter of injected gold nanoparticles. In total, there are five combination models: model 1 consists of tumor category, the diameter of injected gold nanoparticles, TIC curve parameters, mean SWE value, CT radscore, BMUS radscore, and CEUS radscore; model 2 consists of tumor category, mean SWE value, CT radscore, BMUS radscore, and CEUS radscore; model 3 consists of tumor category CT radscore; model 4 consists of tumor category, mean SWE value, BMUS radscore, and CEUS radscore; model 5 consists of tumor category, tumor volume, and the diameter of injected gold nanoparticles.

The classification model used for tumor stratification is constructed using Logistic Regression (LR) and Support Vector Machine (SVM) models. The performance of classification models were assessed by calculating accuracy, sensitivity, specificity, the area under receiver operating characteristic curve (AUC).

The regression model used for precise prediction of gold nanoparticle accumulation is developed using K-Nearest Neighbors (KNN). The predictive performance of the model is evaluated by R-squared ($R^2$), Mean Squared Error (MSE), Root Mean Squared Error (RMSE), and Mean Absolute Error (MAE).

2.8. Masson's trichrome staining histological analysis.

Three to four tumors from each type were randomly chosen, with one-third of each tumor fixed in a 40% formaldehyde solution for Masson's trichrome staining. Utilizing Image J software, five areas of equal size were randomly selected from each Masson-stained tumor section. The relative area of collagen within these sections was calculated and averaged to estimate the collagen content across the entire tumor.

2.9. Statistical analysis

The statistical software used in this study was SPSS 25.0. Measures of central tendency and dispersion were expressed as median and interquartile ranges (IQR; 1st–3rd quartile). Univariable comparisons were performed with the use of the independent-samples t test or Mann-Whitney U test, Kruskal-Wallis test, and chi-square test for normally distributed, nonnormally distributed, and categorical variables, respectively, and a p-value of less than 0.05 indicates a statistically significant difference. The p-value correction method for pairwise comparison within the group was Bonferroni correction. The DeLong test is used to compare whether there are significant differences in the area under the curve (AUC) of independent receiver operating characteristic (ROC) curves.

3. Results

3.1. Characterization of Gold Nanoparticles and PEGylated Gold Nanoparticles

Utilizing the Turkovich-Frens method and seed growth method, we synthesized a range of GNPs comprised of three size series (15nm, 40nm, 70nm), as depicted in TEM images and corresponding diameter distributions in Figure 1a. The diameters of GNPs in TEM images were $13.67 \pm 1.11$ nm, $43.80 \pm 4.81$ nm, and $76.67 \pm 5.11$ nm, respectively. The hydrodynamic diameters of GNPs characterized closely mirrored the core diameters ($15.87 \pm 0.45$ nm, $47.06 \pm 0.74$ nm, $76.36 \pm 0.42$ nm). The zeta potentials of GNPs were measured to be $-27.49 \pm 2.85$ mv, $-27.56 \pm 2.11$ mv, and $-22.66 \pm 2.40$ mv. Then gold nanoparticles were coated with polyethylene glycol (PEG) to ensure their stability in vivo. After PEGylated, the zeta potentials of PEG-GNPs are about electrically neutral (Figure 1c). Dynamic light scattering (DLS) shows an increase of approximately 10-20 nm (Figure 1d), and UV-Vis spectroscopy will redshift about 3 nm (Figure 1b). These characterizations indicate successful conjugation of PEG with the GNPs.

3.2. Tumor Biological Characteristics and Imaging Heterogeneity

To explore the differences between tumors and individuals, we established seven models of subcutaneously transplanted tumors. This array encompassed a wide range of cancer types: B16 (Melanoma, n = 28), E0771 (Breast Cancer, n = 24), Hepa1-6 (Hepatocarcinoma, n = 22), LLC (Lewis Lung Carcinoma, n = 28), MB49 (Bladder Carcinoma, n = 27), MC38 (Colon Carcinoma, n = 34), PAN02 (Pancreatic Adenocarcinoma, n=20). The final analysis included a cohort of 183 mice. The dispersions of tumor volumes were assessed by the Quartile Coefficient of Dispersion (QCD). Significant variabilities were observed not only in tumors ($p < 0.000$) but also in individuals. The overall median tumor volume was 385.17 mm³ [222.78mm³-520.19mm³], with a QCD of 0.41, indicating substantial heterogeneity. Especially within the Hepa1-6 hepatocarcinoma model, the tumor volume showed a QCD of 0.43 and a median of 300.76 mm³ [190.83mm³-473.59mm³].

To delve deeper into the heterogeneity of tumors and individuals, computed tomography (CT) and ultrasound (US) imaging were employed for each tumor. Despite the minimal dispersion in the mean values for both CT and SWE ($QCD_{CT-mean}$ 0.1, $QCD_{SWE-mean}$ 0.1), variations in the average CT and SWE mean values among different tumors were observed (both $p < 0.001$). This indicates that even within a similar range of CT and SWE mean values, the radiological characteristics of tumors can vary due to their unique biological attributes. In the Time-Intensity Curve (TIC) analysis, a larger proportion of tumors will display wash-out curve (110 of 183 [60%] vs 73 of 183 [40%]). Specifically, B16 tumors predominantly showed a washout pattern (23 of 28 [82%]), whereas E0771 tumors were more commonly plateau type (16 of 24 [67%]). Additionally, quantitative parameters derived from the TIC, such as Wash-in Slope (WIS), Time to Peak (TTP), Peak Intensity (PI), and Area Under the Curve (AUC), also demonstrated notable variations among different tumor types (all $p < 0.05$). We calculated the QCD for each quantitative parameter, uncovering that even within the same type of tumor. The hemodynamic characteristics of each tumor exhibited significant variability.

### 3.3. Heterogeneity of Gold Nanoparticles accumulation in Tumors

The accumulation of Gold Nanoparticles across diverse tumor types was quantified as a percentage of the injected dose per gram of tissue (% ID/g). The median GNP accumulation of all tumors was 3.49% ID/g (2.30%ID/g - 6.27% ID/g, QCD = 0.42). This finding is in concordance with the results of a recent meta-analysis conducted by Lin et al. (2023), which reported a median DE of 3.45% ID/g. Notably, variations in GNP accumulation were observed among different tumor types ($p < 0.05$). In Hepa1-6 tumors, marked heterogeneity in GNP accumulation was observed among individual tumors (QCD = 0.51). Concerning the influence of particle size on nanoparticle DE, significant disparities were noted in the uptake in LLC and MC38 tumors (all $p <0.05$). However, no significant differences were detected in the uptake of GNPs of varying particle size ($p = 0.295$) without distinguishing cancer types.

### 3.4. Associations between Tumor Biological Characteristics and Imaging Heterogeneity with GNP accumulation in tumors

To further investigate the factors influencing the GNP accumulation in tumors, as well as the association between tumor biological, imaging characteristics and GNP accumulation, tumors were categorized into high uptake and low uptake groups based on the median GNP accumulation in the training set. The median GNP accumulation in the training set was 3.45% ID/g (2.32% ID/g -5.76% ID/g, QCD = 0.43). There were 94 tumors in the low uptake group and 89 tumors in the high uptake group. No difference for tumor biological characteristics and imaging parameters was observed between the training and test set (all $p>0.05$).

In univariable logistic regression analysis, tumor type, tumor volume, and mean SWE value were significantly correlated with tumor accumulation (all $p<0.05$). Multivariable stepwise analyses showed tumor type and mean SWE value were significantly correlated with tumor accumulation (all $p<0.05$).

Spearman correlation analysis revealed that tumor volume was the only variable negatively correlated with tumor accumulation across all tumors ($r = -0.208$, $p < 0.001$). Specifically, in B16 melanoma, both tumor volume ($r = -0.416$, $p = 0.028$) and mean SWE values ($r = -0.539$, $p = 0.003$) showed negative correlations with tumor accumulation. In the Hepa1-6 liver cancer model, both PI ($r = -0.464$, $p = 0.030$) and AUC ($r = -0.457$, $p = 0.033$) were negatively correlated with tumor accumulation. In MB49 bladder cancer, WIS ($r = 0.591$, $p = 0.001$) and mean SWE values ($r = -0.402$, $p = 0.037$) were significantly associated with GNP accumulation. Meanwhile, in MC38 colon cancer, a negative correlation was observed between mean SWE values ($r = -0.393$, $p = 0.022$) and GNP accumulation.

### 3.5 Performance of radiomics models for tumor stratification

In our study, the training set was used to build the model, LR-based models showed better performance than SVM-based models. Among all LR-based models, model 1, which is a composite radiomics model consisting of tumor category, the diameter of injected gold nanoparticles, TIC curve parameters, mean SWE value, CT radscore, BMUS radscore, demonstrated the best predictive performance. In the training set, model 1 had an AUC of 0.88, specificity of 0.79, sensitivity of 0.86, and accuracy of 0.83. In the test set, it showed similar performance, with an AUC of 0.85, specificity of 0.87, sensitivity of 0.79, and accuracy of 0.83. Compared to the baseline model consisting of tumor category and the diameter of injected gold nanoparticles, the classification radiomics model is better at identifying tumors with high gold nanoparticle uptake both in training and test sets(Delong test, both p<0.05). The decision curve analysis indicates that, compared to the baseline model, the composite radiomics model provides greater net benefit when the threshold probability is greater than 0.25.

3.6 Performance of radiomics models for precise prediction of gold nanoparticle accumulation

To further accurately predict the intratumoral accumulation of gold nanoparticles, we constructed a regression prediction radiomics model. Among all KNN-based models, model 1, which combines tumor category, the diameter of injected gold nanoparticles, TIC curve parameters, mean SWE value, CT radscore, BMUS radscore, and CEUS radscore, exhibited the best performance for accurate prediction of intratumoral gold nanoparticle accumulation. In the training set, model 1 achieved an r of 0.800, $R^2$ of 0.638, MAE of 0.013, MSE of 0.000, and RMSE of 0.018. In the test set, model 1 achieved an r of 0.764, $R^2$ of 0.580, MAE of 0.0.015, MSE of 0.000, and RMSE of 0.019.

3.7 Exploratory analysis of model-driven pathological mechanism

In addition, previous univariate and multivariate logistic regression analyses identified the mean shear modulus from SWE as an independent predictor of gold nanoparticle accumulation in tumor tissue. To further elucidate the underlying mechanisms, comparisons and correlation analyses were conducted. Firstly, tumors grouped by high and low gold nanoparticle uptake showed a marked difference in collagen content, with the high uptake group exhibiting significantly less collagen ($P < 0.001$) . Then, a comparison of the average SWE shear modulus with tumor collagen content revealed a statistically significant positive correlation ($r = 0.393$, $P = 0.018$). Finally, analysis disclosed a negative correlation between tumor collagen content and the accumulation of gold nanoparticles ($r = -0.563$, $P < 0.000$), suggesting that tumors with more collagen accumulate fewer nanoparticles.

4. Discussion

Nanoparticle-based nanomedicines exploit their size advantages for enhanced tumor-targeted drug delivery, promising improved efficacy and reduced toxicity. This concept is grounded in Maeda's 1986 finding of increased nanoparticle extravasation and prolonged retention in tumors due to the EPR effect. Despite this theoretical promise, NPs' clinical progress has faced setbacks, often not outperforming small molecule drugs. Maeda (2015) recognized EPR heterogeneity among individuals, causing variable NPs accumulation and diluting their average clinical advantage over small molecules. This variability is influenced by nanoparticle properties, individual variations, and tumor-specific factors.

Numerous studies have highlighted the decisive role of certain nanoparticle traits in dictating tumor accumulation. By manipulating parameters such as size, charge, shape, and targeting ligands, one can exert control over a nanoparticle's tumor accumulation prowess. Gold nanoparticles, distinguished by their exceptional chemical stability and biocompatibility, are amenable to precise quantification in biological tissues through inductively coupled plasma mass spectrometry (ICP-

MS). Thereby, gold nanoparticles were selected as a model nanomaterial in this research. The present study demonstrated statistically significant differences in intratumoral accumulation of gold nanoparticles with varying diameters in LLC and MC38 tumors ($P < 0.05$). However, when considering tumors regardless of their histological origin, no such statistical difference was observed ($P = 0.295$). Possibly, on the one hand, due to the relatively narrow size range (15, 40, and 70 nm) used in this experiment, size might not have produced discernible differences in accumulation performance. On the other hand, this result suggests that nanoparticle size may have a limited impact on tumor accumulation and fails to account for the vast disparities in accumulation within and between different tumor types. These findings resonate with those of Warren Chan and colleagues, who, through meta-analyses, emphasized the critical importance of investigating tumor heterogeneity in determining nanomedicine accumulation.

The heterogeneity of the tumor, encompassing factors like tumor size, vascular permeability, stromal composition, and receptor expression levels, critically governs nanoparticle accumulation. This notion is confirmed by the statistical analysis of the experimental data. The result revealed pronounced inter- and intra-tumoral heterogeneity in both biological and imaging features across different tumor types and individuals. Our team's previous work utilizing fluorescent imaging of mice tumor models, as well as Song Kim's group's immunohistochemical analysis of patient-derived xenograft (PDX) models, collectively attest to the pivotal role of tumor heterogeneity in dictating the differential accumulation of identical nanomedicines across distinct tumors. Unraveling the influence of tumor heterogeneity on nanomedicine accumulation could reveal governing principles and underlying mechanisms. However, these studies relied on ex vivo tissue analysis, precluding non-invasive prediction of accumulation before biopsy and hindering personalized application.

Medical imaging modalities enable non-invasive tumor visualization and quantitative parameter extraction, reflecting tumor dimensions, morphology, vascularity, stromal components, and surrounding tissue properties. CT imaging is the most commonly used imaging modality in clinical radiology, providing information on tumor size, shape, density, and other characteristics. Ultrasound imaging utilizes high-frequency sound waves to produce images of soft tissues and organs, allowing real-time tissue and organ functional status observation. Contrast-enhanced ultrasound (CEUS) can evaluate tumor blood flow and perfusion, while shear wave elastography (SWE) is an extension of ultrasound elastography based on strain imaging, providing a two-dimensional map of tissue stiffness. SWE quantitatively measures tissue stiffness using shear modulus (kPa) and shear wave speed. This technique is primarily used clinically to evaluate liver fibrosis and assess the benign or malignant nature of breast and thyroid lesions, reflecting tissue hardness and fibrosis. This study employed CT and ultrasound multimodal imaging to provide a comprehensive perspective on tumor biology and heterogeneity. Given their routine use and non-invasive nature in clinical oncology, these imaging techniques hold immense potential for predicting nanomedicine tumor accumulation and translational value. The experimental data identified tumor type, mean CT value, and mean SWE shear modulus as independent predictors of intratumoral accumulation. However, when tumor type was disregarded, only tumor volume showed a negative correlation with intratumoral accumulation. In addition, mean CT value and SWE shear modulus displayed correlations only in certain tumors, such as B16 melanoma. These results demonstrate the difficulty in deciphering accumulation patterns in highly heterogeneous tumors using single features.

As a high-throughput computational technique, radiomics extracts numerous quantitative features from imaging data. It holds the potential to capture micro-level changes in cells, proteins, and molecules, thus facilitating personalized treatment planning, disease diagnosis, therapeutic response assessment, and prognosis prediction. Employing machine learning algorithms to analyze this multidimensional data allows for accurate lesion classification and quantification, surpassing the predictive power of single imaging parameters. This study found that, compared to a baseline model excluding imaging data, a multimodal radiomics composite model more effectively classified tumors, achieving an AUC of 0.87 in the test set. Such a result demonstrates the excellence of radiomics models in predicting nanomedicine accumulation in tumors. Model 1 performed better than Model 2, which indirectly suggests the value of blood perfusion in predicting the accumulation of gold nanoparticles in tumors. However, no effective CEUS radiomics features were selected in the classification model. It is possibly attributed to the low resolution of the CEUS images and the extraction of features from only a single image at peak intensity, which limited the value of the radiomics features that could be obtained. To further refine the prediction of gold nanoparticle accumulation in tumors, we developed a regression model. Although the multimodal radiomics composite model achieved a test set $R^2$ of 0.51, it still exhibited promising predictive potential compared to the baseline model.

Wang et al. studied that SWE shear modulus reflects tumor stiffness and tissue pressure, negatively correlating with nanomedicine delivery and positively correlating with collagen content in the tumor. In this study, we analyzed the relationship between relative collagen area, mean SWE shear modulus, and intratumoral accumulation across different tumors and individuals. Results reveal a positive correlation between relative collagen area and mean SWE shear modulus, and a negative correlation with intratumoral accumulation, consistent with previous findings.

These results suggest that conventional imaging methods, coupled with Radiomics and machine learning, can effectively stratify patients, enabling precision therapy. Regrettably, our study has some limitations as follows: among the multitude of CT features, none emerged as significantly influential on model prediction, potentially stemming from CT's lower soft tissue resolution. Furthermore, the clinical-grade CT and ultrasound scan machines used in this study offer lower resolutions than those dedicated to small animal imaging, potentially underestimating tumor and inter-individual heterogeneity. Future endeavors may benefit from incorporating MRI scans to capture additional soft tissue information, integrating multiple modalities (e.g., T1W, T2W, DWI), and employing novel model architectures to enhance prediction accuracy. Additionally, our study included only 183 tumor datasets, which, given the high-dimensional nature of radiomics models, suggests that the sample size should be further expanded for robust analysis.

In summary, a machine learning model built upon multi-modal radiomics derived from CT and ultrasound images serves as a valid approach to predict gold nanoparticle accumulation in tumors. Integrating tumor biological and imaging features into a multi-modal radiomics model enables tumor stratification and prediction of nanoparticle deposition. This work contributes to the evolving understanding of how tumor heterogeneity shapes nanomedicine accumulation and underscores the promise of harnessing advanced imaging techniques and machine learning algorithms to inform personalized nanotherapeutic strategies.

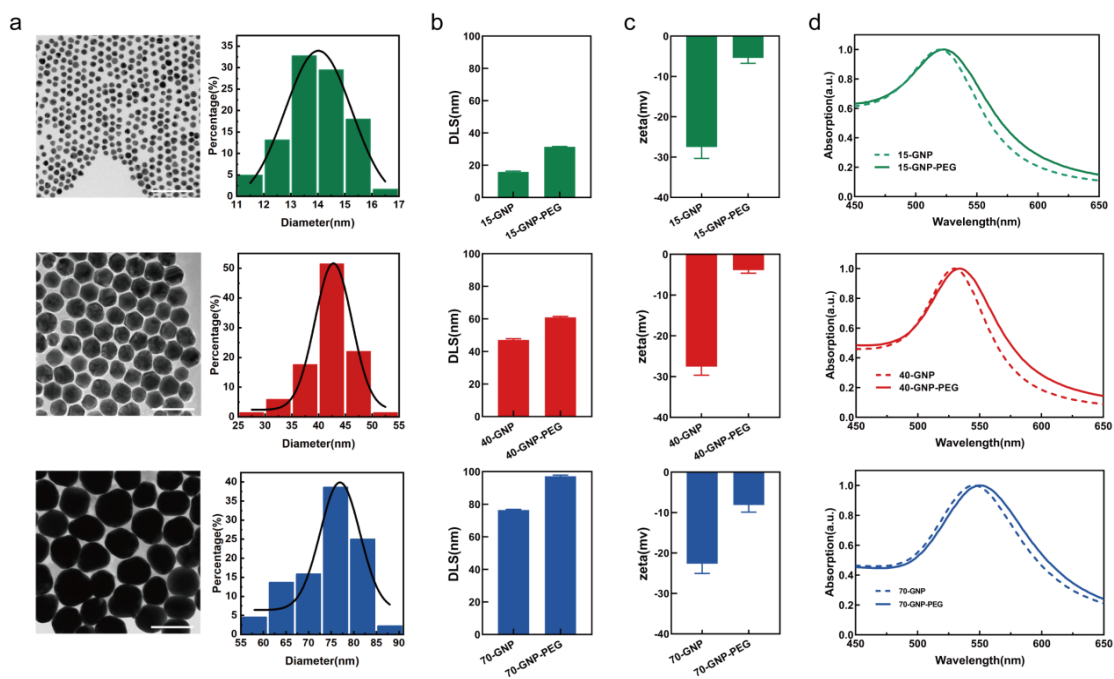

Figure 1 (a) TEM images of GNPs with three different particle sizes (bars: 100 nm) and particle size distribution histogram. (b.c) DLS and Zeta potentials of 15 nm, 40 nm, 70 nm GNP before and after PEG modification. (d) UV–Vis absorption spectrum of 15 nm, 40 nm, 70 nm GNP before and after PEG modification.

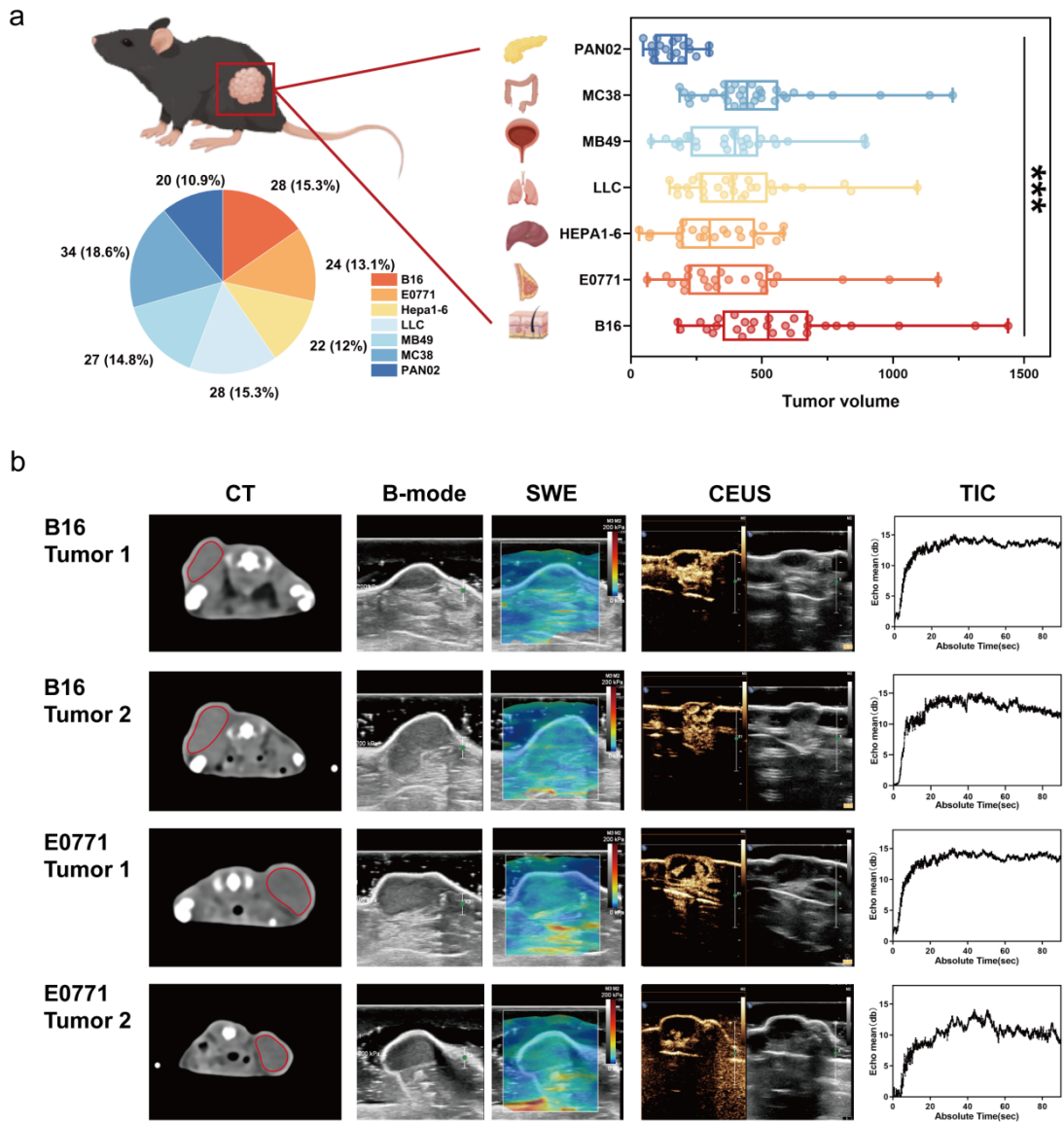

Figure2. The imaging of tumors in the panel of animals.

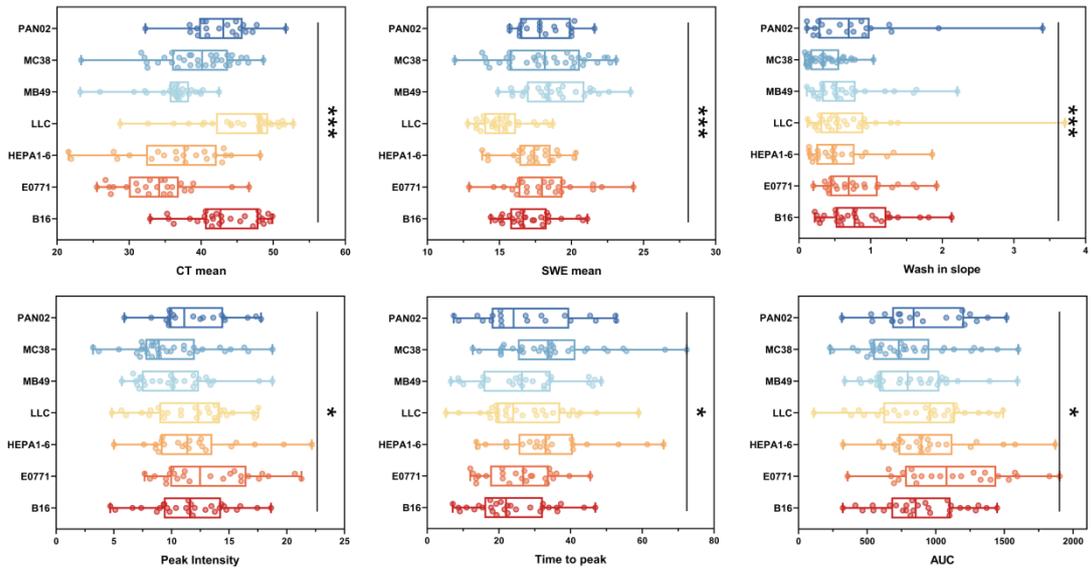

Figure3. The imaging characters of tumors.

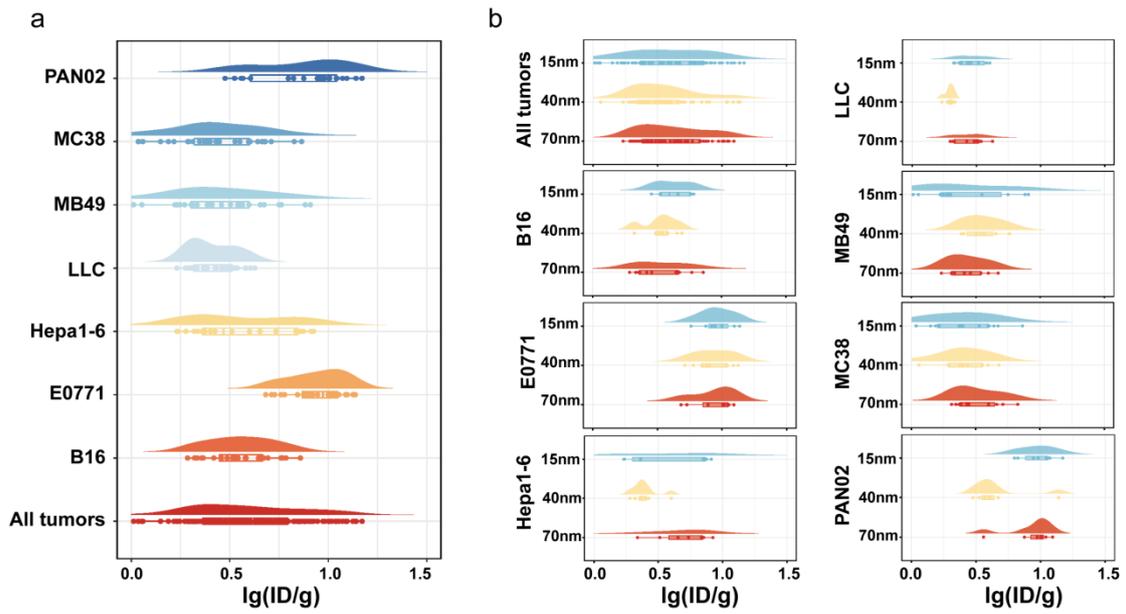

Figure 4. The accumulation of GNPs in different tumors.

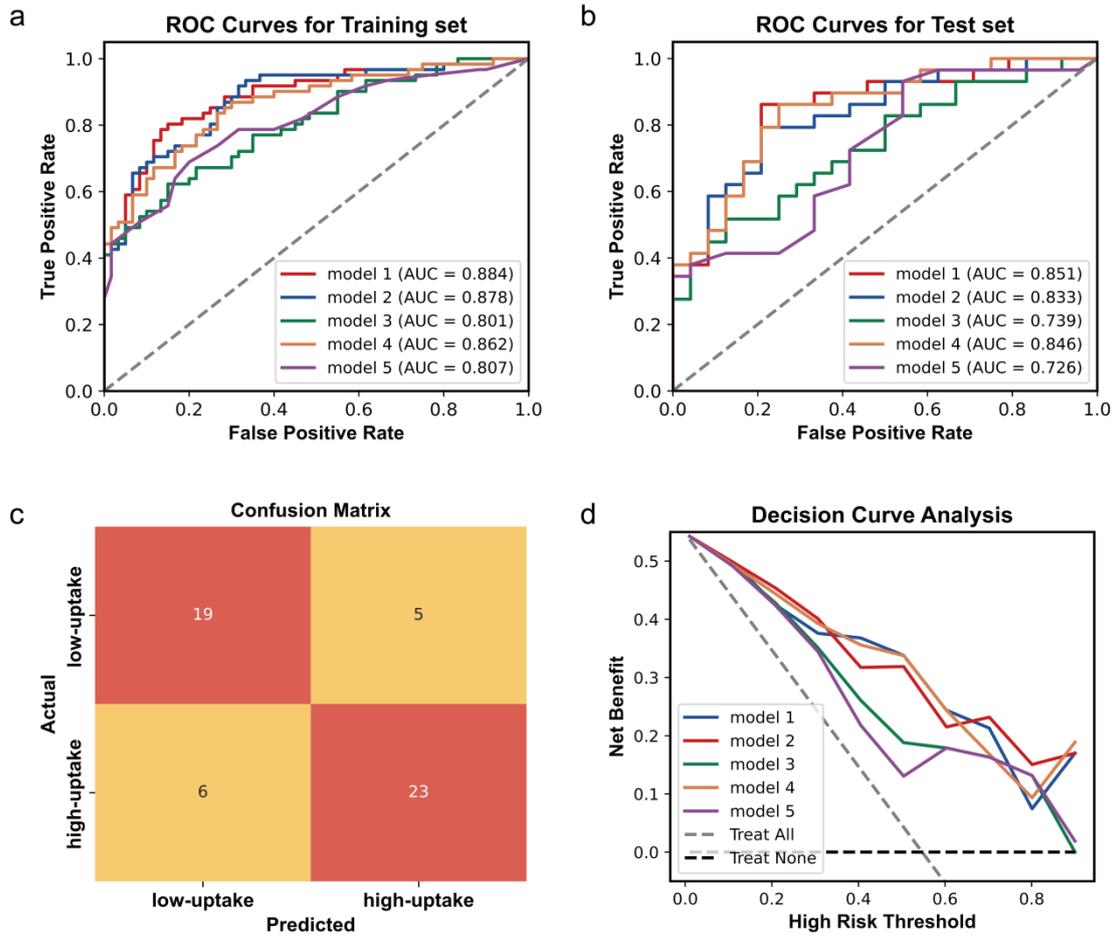

Figure 5 Performance of the models for tumor stratification.

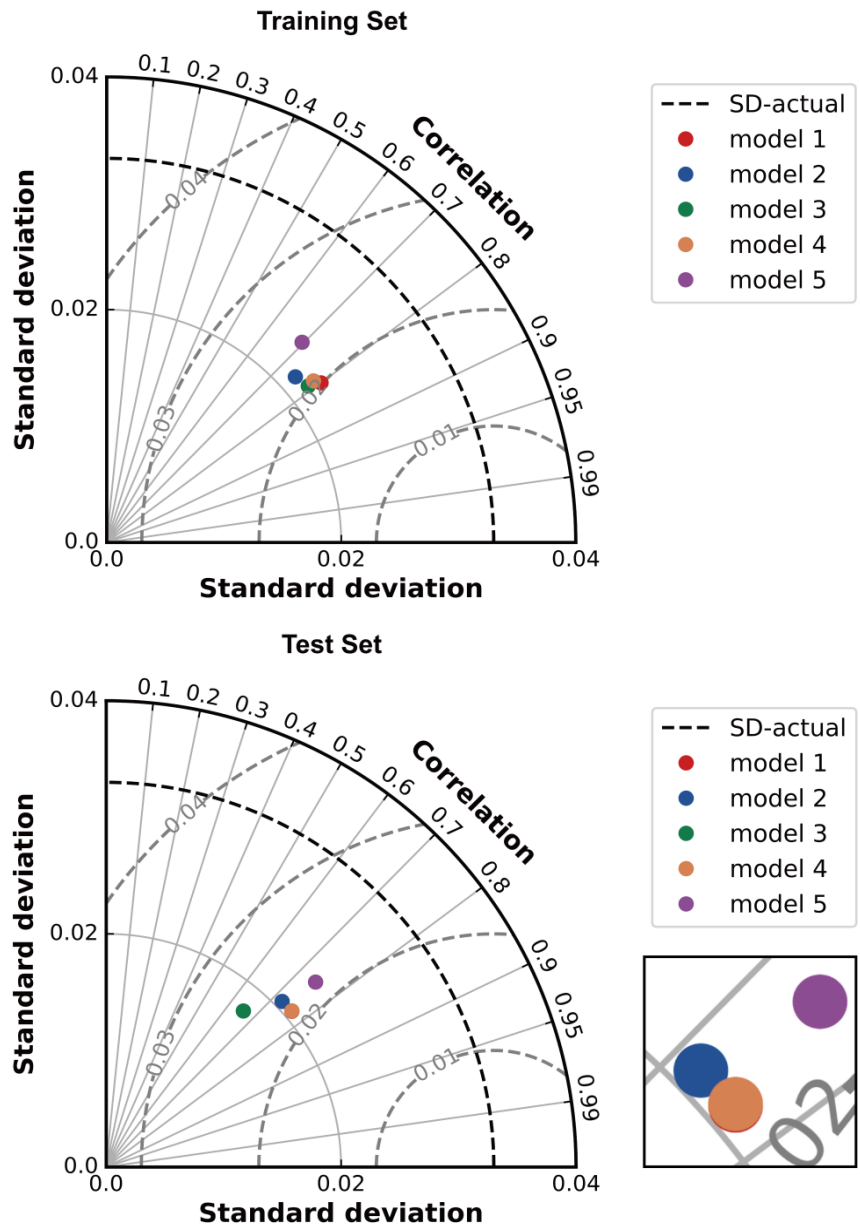

Figure 6 Performance of the models for precise prediction of gold nanoparticle accumulation.

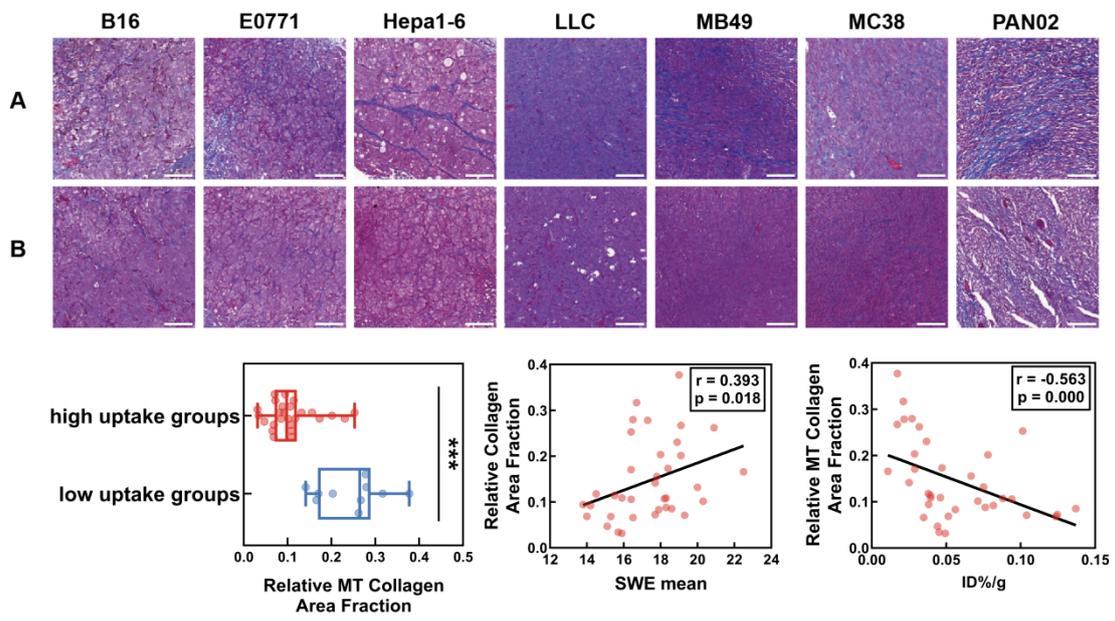

Figure 7. Relationship between MT collagen and SWE mean.